\documentstyle[aps,prl,epsf,epsfig,floats,twocolumn]{revtex}

\begin{document}
\draft
\twocolumn[\hsize\textwidth\columnwidth\hsize\csname @twocolumnfalse\endcsname

\title{Compressibility Divergence and  Mott Endpoints}
\author{G. Kotliar$^{\ast }$ Sahana Murthy$^{\ast }$  
and M.J. Rozenberg$^{+}$}

\address{ Serin Physics Laboratory, Rutgers University, 136
Frelinghuysen Road, Piscataway, New Jersey 08854, USA}

\address{ $^{+}$Departamento de F\'{\i}sica, FCEN, Universidad de
Buenos Aires,
Ciudad Universitaria Pab.I, (1428) Buenos Aires, Argentina.}

\date{\today}
\maketitle
\begin{abstract}

In the context of the dynamical mean field theory of the Hubbard
model, we study the behavior of the compressibility
near the density driven  Mott transition at finite temperatures.
We demonstrate this divergence  using dynamical mean
field theory and Quantum Monte Carlo simulations
in the one band and the two band Hubbard model.
We supplement this result with considerations based on
the Landau theory framework, and discuss the relevance of our
results to the $\alpha$-$\gamma$ endpoint in Cerium.
\end{abstract}

] 

The Mott transition, namely the metal-insulator transition (MIT) driven by
electron-electron interactions \cite{Mott}, is a fascinating phenomenon
realized experimentally in many compounds such as ${\rm V}_{2}{\rm O}_{3}$
and Ni(Se,S)$_{2}$ \cite{Imada:1998}. The Mott transition concept
is also relevant to elements in the lanthanide and actinide series
\cite{Johansson}. Viewed from a broader perspective, the Mott transition
problem forces us to develop tools to describe
materials where the electron is not fully described by either a real space
picture or a momentum space picture, and continues to spur advances in 
many body and electronic structure methods.
 
On the theory side, the Hubbard model is the simplest Hamiltonian that
captures some of the essential physics of the transition.
It has been intensively studied in one dimension, but in this limit no
finite temperature phase transitions can take place.  In recent years, 
theoretical progress has been made in the understanding of the Mott-Hubbard
transition using the dynamical mean-field theory (DMFT) \cite{Georges:1996}.
In this framework,  Mott transition points can be viewed  as 
bifurcation points of a functional \cite{landau,chitra}
of the local Greens function G, or of its associated variable, the Weiss field
which describes the local environment of a correlated site.

The case of the correlation strength $(U)$
driven MIT at half filling, is now well understood.
At temperature $T=0$ there are two bifurcation points, one denoted by 
$U_{c1}(T=0)$ where the insulating solution disappears, and the other
denoted by $U_{c2}(T=0)$ where the metallic state disappears 
in a fashion reminiscent to the Brinkman Rice scenario \cite{brinkman}. 
It was found that in the $U$-$T$ phase diagram of the frustrated Hubbard model
there is a first-order MIT line \cite{krauth:1993,rzk:1994}
that ends in a finite temperature second-order critical point
$(T_{MIT}, U_{MIT})$ which has the character of a regular Ising bifurcation
with a rapid variation of the susceptibility connected to the double
occupancy \cite{Rozenberg:1999,Kotliar:2000}. At higher temperatures the
$U_{c2}(T)$ and $U_{c1}(T)$ lines become crossover lines, 
which have a well defined
experimental significance discussed in 
\cite{rozenberg:1995}.

The zero temperature aspects of the doping driven MIT were
studied in \cite{Fisher:1995}. It was shown that there are two
solutions in an area bound by the curves ${\mu }_{c1}$, where the insulating
solution disappears, and ${\mu }_{c2}$, where the metallic state disappears.
The  finite temperature aspects of the doping driven Mott transition
have not been investigated so far. This is the subject of this paper. 
We will not consider the effects of long range order.

Our main interest is the behavior of the charge compressibility near the 
doping driven Mott transition in the paramagnetic phase at finite 
temperatures.
Furukawa and Imada \cite{Furukawa} pointed out that the compressibility
diverges at the density driven Mott transition in 2-dimensions at $T=0$.
This behavior has been also observed on other models of correlated electron
systems such as the t-J model indicating that this phenomenon is quite 
general \cite{kohno}. Simple models of the Mott transition, 
such as the Gutzwiller approximation or the slave-boson
approach predict a finite compressibility \cite{gutzwiller}.
It is very important to understand the physical origin of this result, 
and to see if it is realized in the DMFT solution of the Hubbard model.
The previously  investigated bifurcation points
within the DMFT, have either a finite charge compressibility, such as
in the $T=0$ density driven Mott transition, or a vanishing charge
compressibility, as in the $T=0$ correlation driven transition. 

Our study of the neighborhood of the Mott transition endpoint is
relevant to materials which have a finite temperature isostructural
phase transition  such as Cerium. This can be seen by generalizing 
the derivation of the Landau free energy of \cite{Kotliar:2000}. 
Near the transition, the free energy 
of more complicated models would have the same form as that studied in
our paper.

We start with the m-band degenerate Hubbard model: 
\begin{eqnarray}
H-\mu N &=&-\frac{t}{\sqrt{z}}\sum_{\langle ij\rangle m\sigma }c_{im\sigma
}^{\dagger }{c_{jm\sigma }}+{\frac{U}{2}}\sum_{i,m,m^{\prime }\sigma }%
\hspace{-0.25cm}n_{im\sigma }n_{im^{\prime }-\sigma }  \nonumber \\
&&+{\frac{U}{2}}\sum_{i,m\neq m^{\prime }\sigma }{\ \hspace{-0.25cm}%
n_{im\sigma }}n_{im^{\prime }\sigma }-\mu \sum_{im\sigma }n_{im\sigma }
\label{eqn:hubbard}
\end{eqnarray}

The first term describes the hopping between nearest neighbors $\langle
ij\rangle $ on a lattice with coordination number $z$. $m, m^{\prime}=1,2$
are the band indices and $\sigma = \uparrow,\downarrow$ labels the spin
index. The parameter $U$ is the energy cost associated with having a double
occupancy on each site. $t$ and $U$ are assumed to be independent of the
band indices and we concentrate on the paramagnetic phase. 
In the limit of infinite dimensions, $z\rightarrow\infty$, 
this model can be mapped onto a
single-impurity Anderson model (SIAM) supplemented by a self-consistency
condition \cite{Georges:1996}. The DMFT equation of the model reads 
\begin{equation}
t^{2}G_{m\sigma}(i{\omega}_n)[\Delta,\alpha]=\Delta(i{\omega}_n)
\label{eqn:DMFT}
\end{equation}
where ${\omega}_n$ are the fermionic Matsubara frequencies, $G_{m\sigma}$ is
the impurity Green's function and $\Delta$ is the hybridization function of
the SIAM. Here, $\alpha$ denotes the control parameter such as temperature $%
T $, Coulomb repulsion $U$, chemical potential $\mu$, etc. We adopt a
semicircular density of states, ${\rho}(\epsilon) = ({\frac {2} {\pi D} } ) 
\sqrt{1 - {({\epsilon/D})^2}} $ where the half-bandwidth $D=2t=1$ is our
unit of energy.

\begin{figure}[h!]
\begin{center}
\epsfig{file=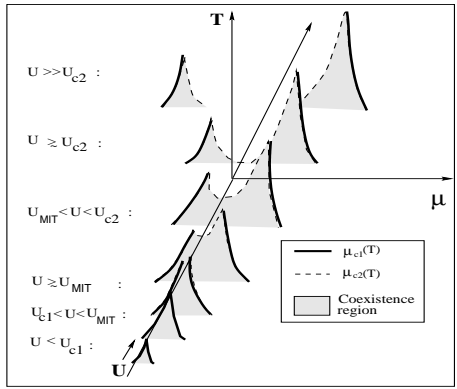, width=6cm,height=5.5cm} \\
\vspace{-0.2cm}
\caption{ Schematic phase diagram for the degenerate Hubbard model. The
cross sections shown are on the $T$-$\protect\mu$ plane for different values
of $U$. ${\protect\mu}_{c1}$ (the heavy line) and $U_{c1}$ are the chemical
potential and interaction respectively at which the insulating solution gets
destroyed. ${\protect\mu}_{c2}$ (the dotted line) and $U_{c2}$ are the
chemical potential and interaction at which the metallic solution gets
destroyed. $U_{MIT}$ is the value of the interaction at which the MIT 
endpoint takes place. }
\label{fig:phased}
\end{center}	
\end{figure}

We now propose a schematic phase diagram (Fig.~\ref{fig:phased}) for the
degenerate Hubbard model. The figure shows cross sections of the $T$-$\mu$
plane for various values of the interaction $U$. In the particle-hole
symmetric case in the 1-band model the peaks have equal height.
The shaded regions indicate coexistence between
metallic and insulating solutions.
The presence  of a coexistence region and a first order phase transition
in 2-band model was apparent in earlier Monte Carlo calculations 
\cite{rozenberg:1997}. 
This also follows from the general form of the Landau functional of this
problem \cite{landau}, which has the same form for the 1-band and 
and the multiband situations.
The finite temperature and general chemical potential aspects of this
problem had not been discussed before in the literature.

\begin{figure}[t!]
\epsfxsize=2.5 in
\epsffile{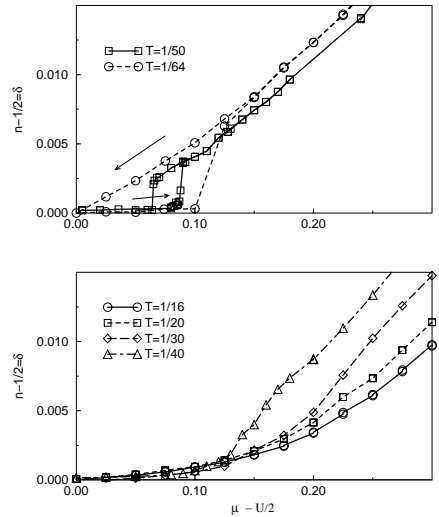}
\vspace{-0.25cm}
\caption{Particle occupation  $\delta=\langle n \rangle$-$1/2$  
per spin as a function of the shifted chemical potential 
$\mu-U/2$ for different temperatures $T$. 
The lower panel shows QMC data obtained at 
$T = 1/40, 1/30, 1/20, 1/16$ (right to left) which are above 
$T_{\mbox{\scriptsize MIT}}$.
The upper panel shows similar data at temperatures
$T=1/64, 1/50$ which are below  $T_{\mbox{\scriptsize MIT}}$. 
The left arrow indicates that starting at high doping, a metallic 
solution can be followed down to half-filling. 
Also, starting from an insulating state at half-filling, 
one can follow the coexisting solution (right arrow) 
by increasing $\mu$ up to a sudden jump in $\delta$.}
\label{fig:nmu1b}
\end{figure}

On solving the DMFT equation (\ref{eqn:DMFT}) iteratively using quantum
Monte Carlo (QMC) methods, we find that at integer filling, for smaller
values of $U$ ($<U_{c1}$ in Fig.~\ref{fig:phased}) the solutions show
metallic behavior while for large values of $U$ ($>U_{c2}$ in Fig.~\ref
{fig:phased}) the solutions show insulating behavior. In Fig.\ref{fig:nmu1b}
we show the results for the doping (per spin) $\delta = \langle n
\rangle-1/2 $ as a function of the chemical potential $\mu$ obtained from
QMC calculations with $\Delta\tau=0.5$ in the single band case. 
The interaction value $U=2.46$  is inside the small region of
coexisting solutions when the system is at half-filling 
\cite{Rozenberg:1999}. 
In the lower panel we show the results at several temperatures above the
critical temperature $T_{MIT}$. As the temperature is decreased the curves 
develop a sigmoidal shape, 
which is a hallmark of the approach to a second order critical point in
Landau theory, as in the familiar Ising mean-field model.

\begin{figure}[t]
\begin{center}
\epsfig{file=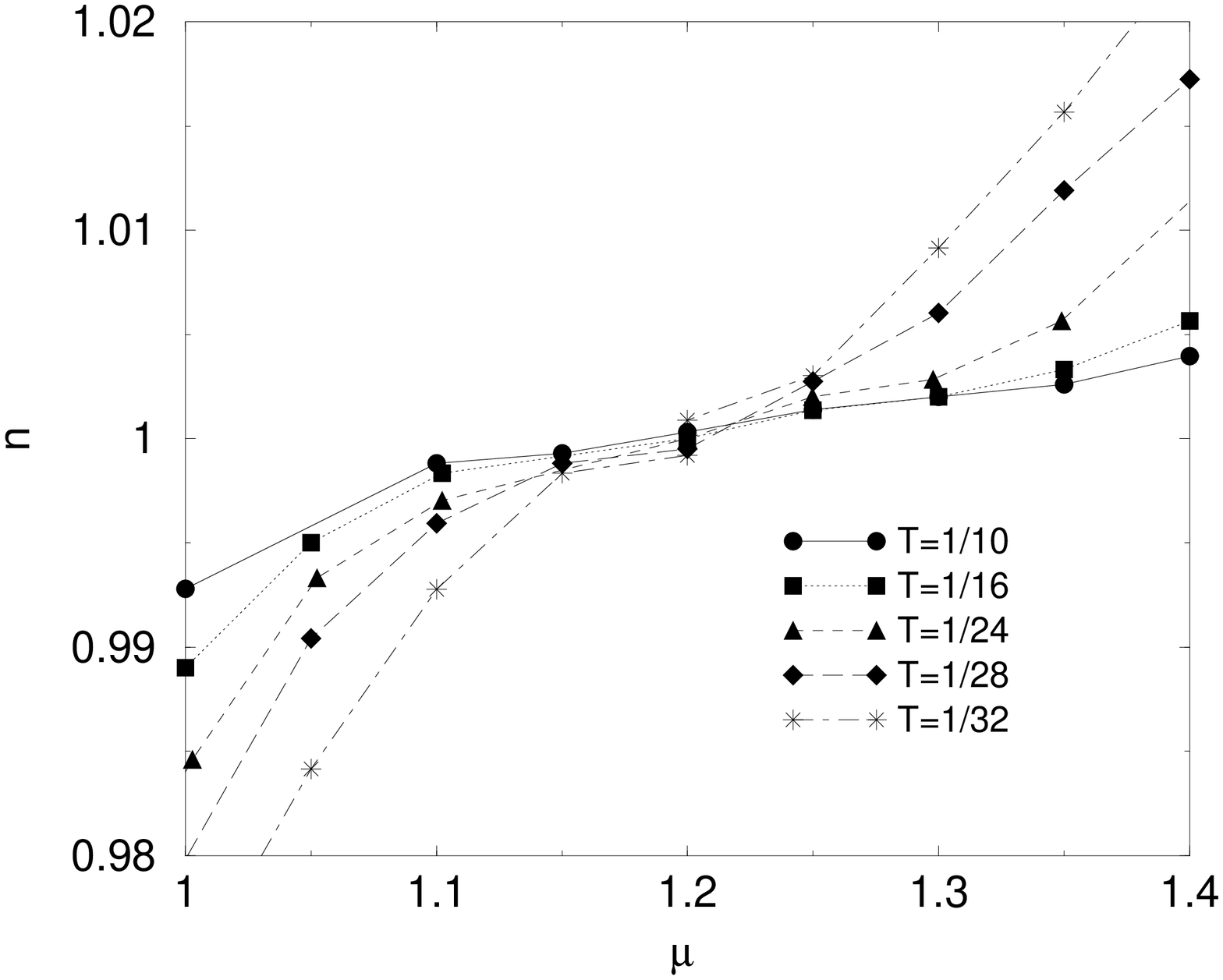,width=4cm, height=5.5cm,angle=-90}
\epsfig{file=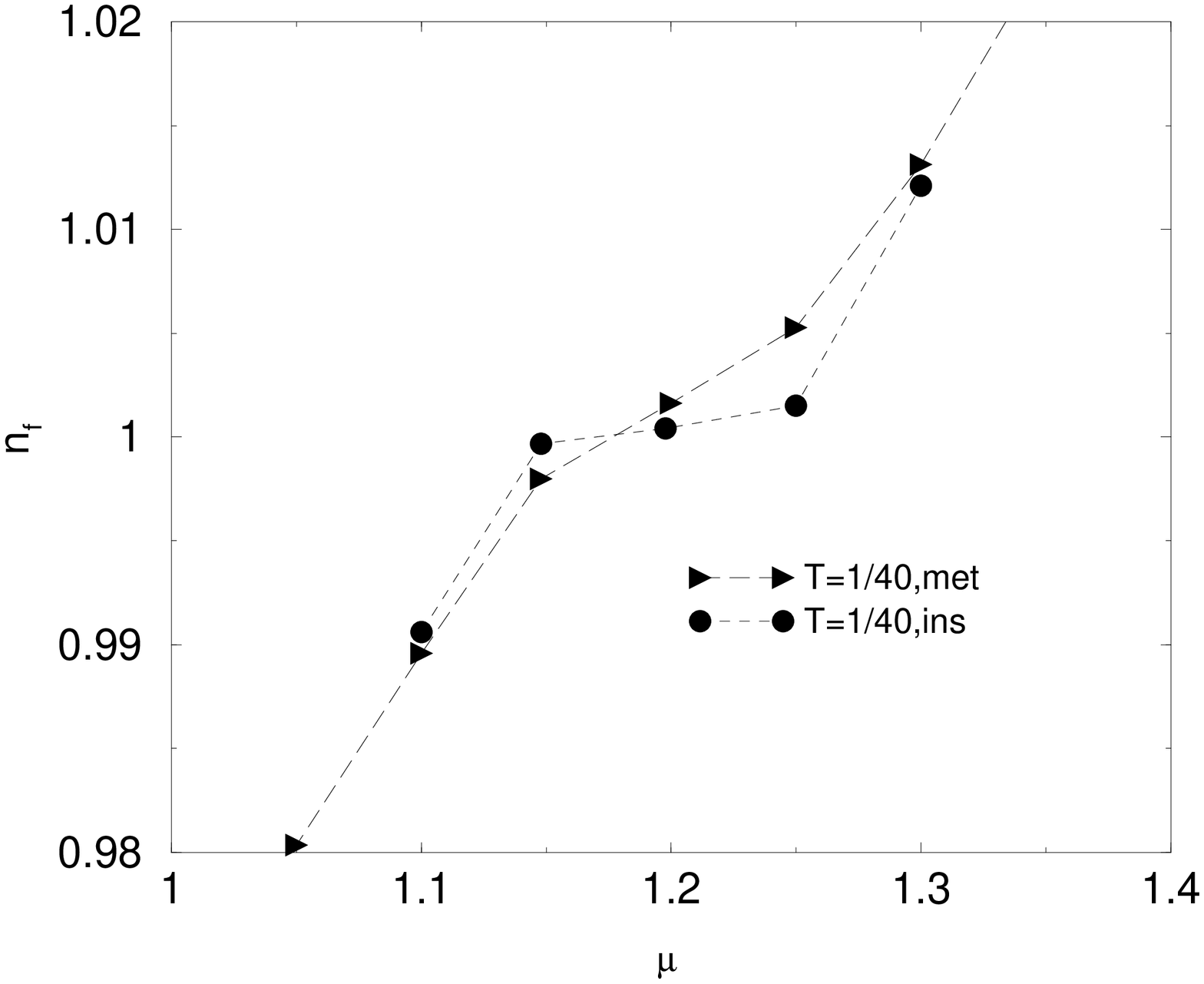,width=3.5cm,height=5.5cm,angle=-90}
\caption{
Particle occupation  $n$ function of $\mu$ for different temperatures. 
in the 2-band model at U=3.0. In the upper panel, the temperatures are
above the critical temperature while in the lower panel, we see
coexistence at $T<T_{MIT}$.}
\label{fig:nmu2b}
\end{center}
\end{figure}

The results for the total occupation number $n$ as a function of $\mu$ in
the 2-band model are shown in Fig.\ref{fig:nmu2b}. The QMC calculation was
carried out at $\Delta\tau=0.25$ and $U=3.0$. At  higher
temperatures, only one solution is present, whereas
on lowering the temperature to $T=1/40$ both metallic and insulating
solutions coexist.

To obtain coexisting solutions below $T_{MIT}$
 we first start at high doping in a metallic state as
indicated by the left arrow in the upper panel of Fig.\ref{fig:nmu1b}. For
this, we use a seed from low $U$ ($<U_{c1}$ in Fig.\ref{fig:phased}) and
continuously evolve towards integer filling through the upper branch
solution which always remains metallic with finite compressibility. 
We use a self-consistent solution as new seed for the 
successive calculations at smaller $\mu$. On the
other hand, a second solution is also present at half-filling in Fig.\ref
{fig:nmu1b} (or near $n=1$ in Fig.\ref{fig:nmu2b}) corresponding to an
insulting-like state, obtained by using a seed from large $%
U$ ($>U_{c2}$ in Fig.~\ref{fig:phased}). This state is essentially
incompressible as $n$ almost remains constant while
increasing $\mu$. This can be continuously followed as $\mu$ is increased
until the eventual jump of $n$ towards the unique solution present at the
higher values of $\mu$. This procedure successfully determined the
location of the coexisting region\cite{Joo}.

Locating these parameters in the schematic phase diagram, we find that the
data for the 1-band model below $T_{MIT}$ in Fig.~\ref{fig:nmu1b}
corresponds to the region near the right peak at $U_{MIT}<U<U_{c2}$ 
in Fig.~\ref{fig:phased}. As we move towards $\delta=0$ in
the metallic phase, we cross the coexistence region and reach the 
${\mu}_{c2} $ line  at $T=1/50$. Upon lowering
$T$ to $1/64$ we can approach $\delta=0$  without crossing the 
$\mu_{c2}$ line. In the 2-band model, it was found that at $T=1/40$ 
the insulating solution disappears as we move away from $n=1$ 
indicating the the ${\mu}%
_{c1} $ line has been crossed indicating that the data in Fig.~\ref
{fig:nmu2b} lies near the right peak at $U_{MIT}<U<U_{c2}$ in Fig.~\ref
{fig:phased}.

The region with two coexisting solutions has two different values of $n$ for
given $T$ and $\mu$. These two coexisting solutions have different free
energies and the actual thermodynamic state of the system will be that of
minimum energy. Therefore, a jump in the particle occupation is predicted at
a first-order line. The actual determination of this line, thus implies a
precise calculation of the free energy which lies outside the scope of the
present work.

\begin{figure}[t]
\begin{center}
\epsfig{file=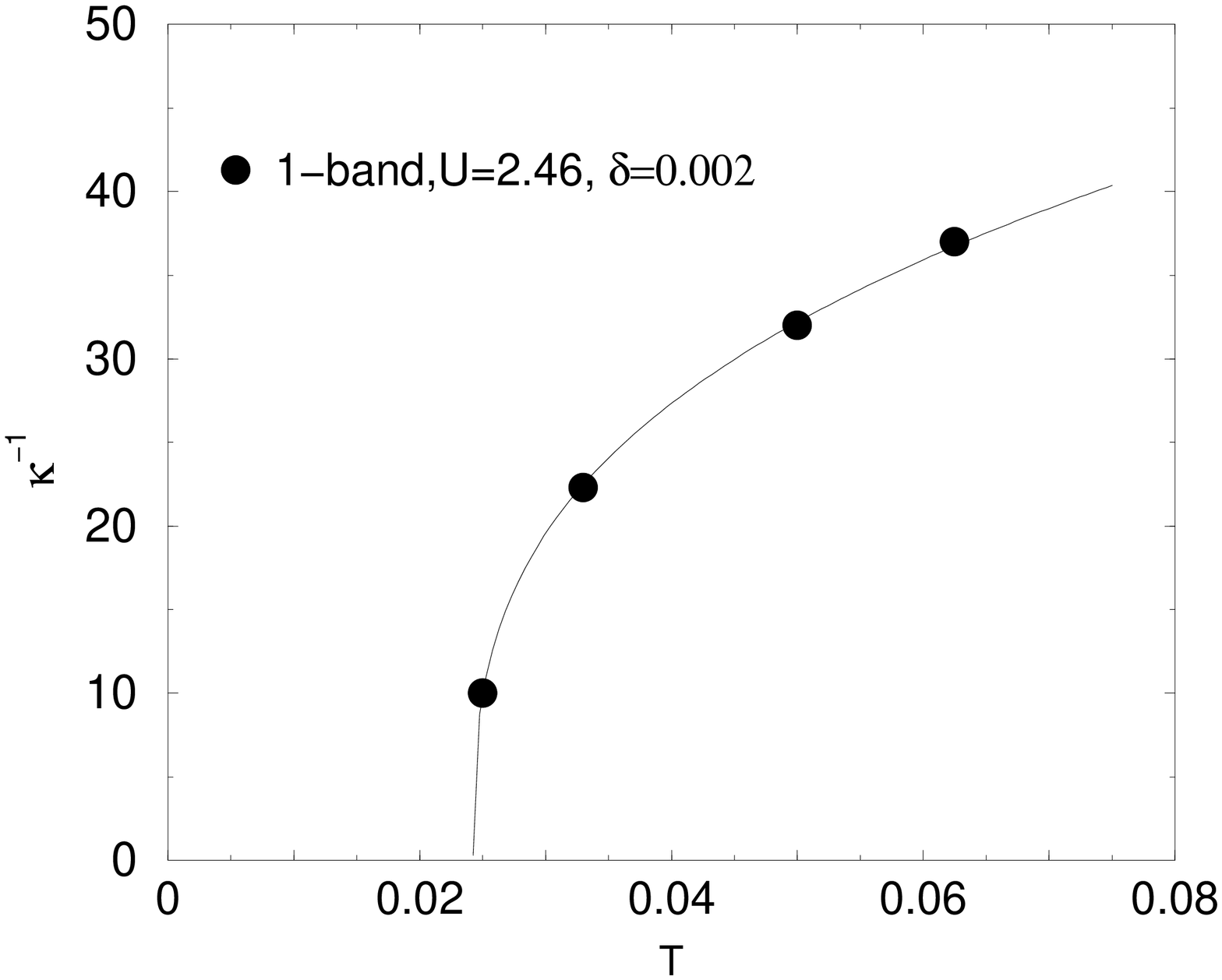, height=6.cm,width=3.0cm,angle=-90} \\
\epsfig{file=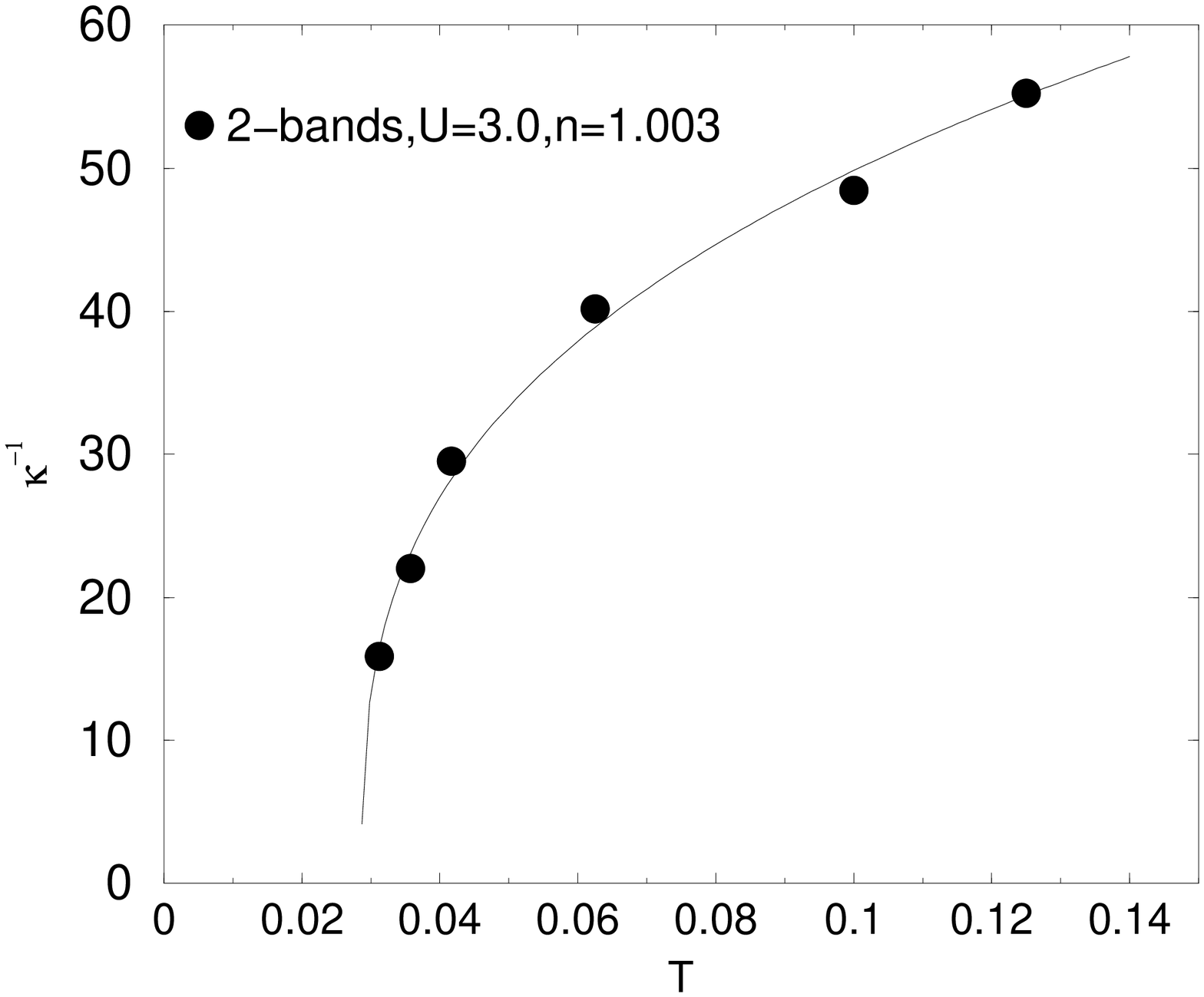, height=5.8cm,width=3.0cm,angle=-90}
\caption{
The inverse compressibility $\kappa^{-1}$, obtained by the numerical 
differentiation of $n$ with respect to $\mu$, at a constant doping 
as a function of $T$. 
Top: 1-band model at $U=2.46$ with doping $\delta=0.002$.
Bottom: 2-band model at $U=3.0$ at  $n=1.003$.
The solid lines are a guide to the eye.}
\label{fig:kappa}
\end{center}
\end{figure}

From the $n$ vs. $\mu$ curves in Figs.~\ref{fig:nmu1b} and \ref{fig:nmu2b} we
computed the numerical derivative of the particle number with respect to the
chemical potential to obtain charge compressibility $\kappa$. 
The results for ${\kappa}^{-1}$ as a function of the temperature 
above $T_{MIT}$ 
(Fig.\ref{fig:kappa}) indicate that ${\kappa}^{-1} \rightarrow 0$ as we pass
through the MIT.

We also like to mention that in our simulations we observed characteristic
effects of enhanced fluctuations and critical slowing down as the MIT
is approached. Hence, simulations have to be done with extreme care, 
appropriately choosing the seeds  for the iterative process 
and gathering many sample points for accurate statistics. 
In practice we use $\sim 10^5$ MC sweeps and run our
calculations in an 8 node parallel cluster.  As we approach the
critical points, we used up to a few hundred iterations to obtain 
converged solutions.

We now complement  our numerical results with 
Landau theory arguments. 
The mean-field equation (\ref{eqn:DMFT}) can be obtained by
differentiating the Landau functional \cite{landau}

\begin{equation}
F_{\mbox{\scriptsize
 	LG}}[\Delta ]=-T\sum_{n}\frac{\Delta (i\omega _{n})^{2}}{t^{2}}+F_{%
\mbox{\scriptsize
imp}}[\Delta ]  \label{f_landau}
\end{equation}
with respect to the hybridization $\Delta (i\omega _{n})$ of the SIAM, which
has the meaning of a Weiss field. $F_{\mbox{\scriptsize imp}}[\Delta ]$ is
the free energy of the SIAM in presence of a hybridization. The Green's
function of the SIAM is $G(i\omega _{n})[\Delta ,\alpha ]=(1/2T)\delta F_{%
\mbox{\scriptsize imp}}/\delta \Delta $. This Landau approach was used to
describe the Mott transition at half filling \cite
{Rozenberg:1999,Kotliar:2000}.

As discussed in Ref.\cite{Kotliar:2000}, the finite temperature Mott
transition is a regular bifurcation point. On differentiating the Landau
functional twice we get a matrix of the form $-{\delta}_{nm}+M_{nm}$ where 
\begin{equation}
M_{nm}=\frac{t^2}{2T}\left.\frac{\delta^2F_{\mbox{\scriptsize
   imp}}[\Delta]}{\delta\Delta(i\omega_n)\delta\Delta(i\omega_m)} \right|_{%
\mbox{\scriptsize cp}}
\end{equation}
acquires a zero mode.

We now make a small change in chemical potential around the critical point
and expand the mean field equation(\ref{eqn:DMFT}) to first order in $%
\delta\alpha=(\mu-\mu_c)$, $\delta\Delta=\Delta(\alpha_c+\delta\alpha)-%
\Delta(\alpha_c)$. We get, 
\begin{equation}
\frac{\delta\Delta (i {\omega}_n)}{\delta\alpha} = \sum_{m} \frac{1}{1-M_{mn}%
} t^2 \frac{ \partial G_{imp}(i {\omega}_n)} {\partial \alpha}
\end{equation}

From (\ref{eqn:DMFT}), the lattice occupation at any site which is identical
to the local impurity occupation is related to the hybridization by $\langle
n \rangle = ({2T}/{t^2}) \sum_n \Delta(i {\omega}_n)$. Thus,

\begin{equation}
\frac{d\langle n \rangle}{d\mu} = 2 \sum_n \left[ \frac{1}{1-M} \right] T 
\frac{\partial G_{imp}(i {\omega}_n)}{\partial \mu}  \label{comp}
\end{equation}

Clearly, unless the derivative in the $rhs$ of (\ref{comp}) is exactly
orthogonal to the zero mode of the matrix M, the bifurcation condition leads
to the singular behavior of the compressibility. 
The QMC studies shown above demonstrate that this orthogonality 
does not occur.
The divergence of the compressibility strengthens the liquid-gas
picture  of the Mott transition presented by Castellani {\it et. al.} 
\cite{castellani}.
  
From the experimental viewpoint, we believe that our results 
highlight important aspects of the
physics of the $\alpha$-$\gamma$ transition in Cerium. The detailed
description  of the non-universal aspects of this material requires 
more quantitative studies or more elaborate models such as those
studied by Held {\it et. al.} \cite{held}.
However, the functional describing these complicated models near 
the finite temperature Mott transition would reduce to the one 
underlying the equations we studied. Hence it is plausible that the 
behavior of the compressibility that we identified in our basic
model applies to the $\alpha$-$\gamma$ transition of Cerium as well.
The divergence of compressibility in
the Cerium $\alpha$-$\gamma$ transition therefore has an electronic
origin and can be understood from model calculations without involving
lattice degrees of freedom (which would renormalize values of the
critical points without changing qualitative features).
In our data, we find  a decrease in compressibility during the 
transition from the insulating to the metallic phase which is 
similar to what has been measured by Beecroft and Swenson in \cite{beecroft}. 
According to the results in \cite{beecroft}, the compressibility 
in the low-pressure $\gamma$ phase (corresponding to the insulating
region) is larger than in the high pressure $\alpha$ phase
(corresponding to the metallic-like state) at low temperatures, 
and this effect becomes less noticeable at higher temperatures. 
Our QMC studies have shown that there is a region of coexistence where 
the metallic phase is extended into the stability region of the 
insulating phase, corresponding to the hysteresis observed in the data in
\cite{beecroft}. 
This was also found in X-ray diffraction studies on Cerium \cite{adams}, where 
the $\gamma$ phase was quenched into into the region of stability 
of the $\alpha$ phase near the phase boundary. 

In summary, we presented a careful QMC numerical solution of the doping
driven Mott transition in the limit of large lattice connectivity. Our study
unveils that the divergence of the compressibility is a generic
feature of the confluence of the finite temperature Mott endpoint. We
understood this divergence in terms of a simple argument based on Landau
theory, which indicates that these results are more general than the
specific models for which the numerical studies were carried out. 
Our results were found to be relevant to the physics of the Cerium 
$\alpha$-$\gamma$ transition.

ACKNOWLEDGMENTS : 
This research was supported by the Division of Materials Research
of the National Science Foundation,  under grant
NSF DMR 0096462, the Division of Basic Energy
Sciences of the Department of Energy  under grant
US\ DOE, grant No.
DE-FG02-99ER45761.
M.J.R.
acknowledges support of Fundaci\'{o}n Antorchas, CONICET (PID $N^{o}4547/96$%
), and ANPCYT (PMT-PICT1855).

\end{document}